\documentclass[showpacs,preprint,aps]{revtex4}
\usepackage{graphicx}% Include figure files
\usepackage{dcolumn}% Align table columns on decimal point
\usepackage{bm}% bold math
\usepackage{amssymb}
\usepackage{amsmath}
\begin{document}
\setcounter{page}{1}
\title 
{Comparing two approaches to Hawking radiation of 
Schwarzschild-de Sitter black holes}
\author
{I. Arraut$^1$, D. Batic$^2$ and M. Nowakowski$^1$}
\affiliation{$^1$  
Departamento de Fisica, Universidad de los Andes, 
Cra.1E No.18A-10, Bogota, Colombia\\
$^2$ Departamento de Matematica, Universidad de los Andes,
Cra 1E, No. 18A-10, Bogota, Colombia}
\begin{abstract}
We study two different ways to analyze the Hawking evaporation of a 
Schwarzschild-de Sitter black hole. The first one uses the standard 
approach of surface gravity evaluated at the possible horizons.
The second method derives its results via the Generalized Uncertainty Principle (GUP) 
which offers a yet 
different method to look at the problem. In the case
of a Schwarzschild black hole it is known that this methods 
affirms the existence of a black hole remnant (minimal mass $M_{\rm min}$)
of the order
of Planck mass $m_{\rm pl}$ and a corresponding maximal temperature
$T_{\rm max}$ also of the order of $m_{\rm pl}$. The standard $T(M)$ dispersion relation
is, in the GUP formulation,  deformed in the vicinity of Planck length $l_{\rm pl}$
which is the smallest value the horizon can take. We generalize the 
uncertainty principle to Schwarzschild-de Sitter spacetime with
the cosmological constant $\varLambda=1/m_\varLambda^2$  and find
a dual relation which, compared to $M_{\rm min}$ and $T_{\rm max}$,
affirms the existence of a maximal mass $M_{\rm max}$
of the order $(m_{\rm pl}/m_\varLambda)m_{\rm pl}$, minimum
temperature $T_{\rm min} \sim m_\varLambda$. As compared to the standard approach
we find a deformed
dispersion relation $T(M)$ close to $l_{\rm pl}$ and 
in addition at the maximally possible horizon
approximately at $r_\varLambda=1/m_\varLambda$. $T(M)$ agrees with the standard results
at $l_{\rm pl} \ll r \ll r_\varLambda$ (or equivalently at $M_{\rm min} \ll M \ll M_{\rm max}$).     
\end{abstract}
\pacs{04.60.Bc, 04.70.Dy} 
\maketitle 

\section{Introduction}
In the recent years, the Generalized Uncertainty Principle (GUP),
the uncertainty relation which includes also gravity effects,  
has gained popularity \cite{adler1, GUP2, GUP3}. 
Especially, in the context of black holes \cite{GUPBH1} and their
evaporation \cite{hawking, adler2, GUPBH2} GUP has proved to be
the harbinger of new, maybe partly also expected 
effects in the context of quantum gravity.
Compared to the standard Hawking radiation
GUP deforms the standard $T(M)$ relation
near the Planck length to the extent that the Planck length becomes
the smallest possible length scale in this context. One can interpret this
result also in a different way: there exist a minimum mass (which is the
black hole remnant) of the order of Planck mass which corresponds to a
maximum temperature, also of the order of Planck mass \cite{adler2, cavaglia}.
This fits neatly 
into the picture of dimensional analysis based on the Newtonian constant 
$G$. It is expected that all scales given by $G$, i.e.
$l_{\rm pl}=1.61 \times 10^{-33} $ cm, $t_{\rm pl}=5.39 \times 10^{-44}$
sec, $m_{\rm pl}=1.22 \times 10^{19}$ GeV and $\rho_{\rm pl}=5.16 \times 
10^{93}$ g ${\rm cm}^{-3}$ (density) are the extreme or limiting 
values which can be attained in a physical situation. It is also expected that
at these scales, special effects of quantum gravity will show up.

Seen from a certain perspective, the early stage of quantum theory
resembles the current state of art of what we call quantum gravity. With
respect to the former, the important 
harbingers of the (those days, new) quantum theory 
were Planck's black body radiation formula and the uncertainty relation 
$\Delta x \Delta p \ge 1/2$ derived in the early days
without the help of Schwarz inequality. Today quantum gravity seems
to offer a very similar state of affairs which, of course, does not imply that
there do not already exist aspirants for a complete quantum gravity theory.
Hawking's theory of black hole evaporation \cite{hawking} is not only a quantum mechanical
effect, but the radiation of black holes is also a perfect black body
radiation. Secondly, the above mentioned Generalized Uncertainty Principle, which
includes gravity effects, has been derived in different contexts: string theory 
and
non-commutative quantum theory. Recently, a simpler derivation of this
uncertainty relation has been found which agrees fully with the 
previous findings \cite{adler1, adler2}.

Since GUP offers a robust tool to probe into quantum mechanics of black holes, 
it is interesting to raise the question, what will actually 
happen if another constant enters the Einstein's equations.
This is not a remote possibility as the recently discovered acceleration
of the universe \cite{accel} cannot be explained without  altering either
the Einstein tensor $G_{\mu \nu}$ or the energy-momentum tensor
for cosmology (including the equation of state). Opting for the
first possibility, any new constant in $G_{\mu \nu}$ is 
independent of the situation to which we apply the
Einstein's equations and therefore a new constant of gravity.
It is worth pointing out that the evidence for the need to change the standard gravity is growing.
Observation of standard candles like type Ia Supernova \cite{riess} and other
key observations in relation with Baryon Acoustic Oscillations \cite{eisenstein},
Cosmic Microwave Background radiation \cite{spergel}, Large 
Scale Structure \cite{Knop2003} and weak lensing \cite{chae} led us to the conclusion 
that the expansion of the Universe as compared to the standard Friedmann 
model is 
accelerated. 
All evidence is in agreement with a positive cosmological constant. Notably, 
the observations seem to favor the equation of state $p=-\rho$ which comes along with
the gravity theory including the cosmological constant $\varLambda$.

If the positive cosmological constant $\varLambda$ explains the recently
discovered accelerated stage of the universe, this constant is, 
beside the Newtonian constant $G=m_{\rm pl}^{-2}$, the
second constant of gravity. It is legitimate to put forth the
question: how does the mass scale $m_{\varLambda}=\sqrt{\Lambda} \ll m_{\rm pl}$
and length scale $r_\varLambda=1/m_\varLambda \gg  l_{\rm pl}$
alter our expectations for quantum gravity. 
In the present paper we first find the Hawking approach to
Schwarzschild-de Sitter black hole radiation. In the second step we elaborate on the
evaporation of this black hole  
utilizing the Generalized Uncertainty Principle (GUP) with $\varLambda$ 
and confirm the results by black body radiation.
We compare the $T(M)$ dispersion relation which we derive from GUP including $\varLambda$ with the
standard expression obtained from surface gravity calculated at an event horizon.
They both agree for intermediate mass range, i.e., masses much bigger
than Planck mass, but much smaller than $M_{\rm max} \sim (m_{\rm pl}/m_{\varLambda})m_{\rm pl}$.
This is what one would expect from GUP which now deforms the standard $T(M)$ relations
at $l_{\rm pl}$ (corresponding to the mass of black hole remnant 
$M_{\rm min} \sim m_{\rm pl}$) and at $r_\varLambda$ (corresponding to $M_{\rm max}$).
A careful analysis performed in this paper reveals 
the following picture: at masses 
close to Planck mass $T(M)$ follows
the behavior found in \cite{adler2} (here $\varLambda$ does not play any significant role),
this is taken over by the standard Hawking, i.e., $T(M)=m_{\rm pl}^2/(8\pi M)$.
As $M$ becomes bigger, the effects of $\varLambda$ become more important. They can still be described 
by the standard approach, i.e., calculating the surface gravity at a horizon where
$\varLambda$ enters now explicitly the expression for $T(M)$. For even higher masses
GUP modifies this standard picture to the extent that there exists a maximum
mass of the order $M_{\rm max}$ beyond which no positive definite solutions exist for $T$.
This means that we have a minimum temperature $T_{\rm min}=T(M_{\rm max}) \sim m_\varLambda$.
In short, GUP results into the existence of a maximum temperature corresponding to a
minimum mass and a minimum temperature corresponding to a maximum mass. 
The latter results is due to $\varLambda$.
We can replace
the mass by length in which case we have a minimum (mentioned already above) and
maximum length. The latter is $r_\varLambda$.

The outline of the paper is as follows. In section 2 we will
determine the full and approximated expressions for the two horizons in the
Schwarzschild- de Sitter case. Section 3 is devoted to the standard
treatment of Hawking radiation of Schwarzschild-de Sitter black holes via the
the surface gravity calculated at the horizons. We will show that only
the first horizon gives physically viable results as $T(M)$ calculated
at the second horizon violates the condition $\partial T/\partial M < 0$.
Section 4 contains the discussion of GUP applied to the black hole
evaporation. We briefly touch the case $\varLambda=0$ to 
show explicitly the major steps involved in the derivation. Then
the generalization to $\varLambda$ will be transparent.
We apply the uncertainty relation with $\varLambda$ to the black hole 
evaporation and find that, as far as the order of magnitude is concerned,
for intermediate masses it agrees with results derived in section 2.
We show the existence of $T_{\rm min}$ and $M_{\rm max}$.
In section 5 we confirm the results obtained in 4 by yet different methods.
Section 6 discusses a different effect of $\varLambda$ in a temperature
perceived at a distance. In section 7 we summarize our findings.
The two appendices are included for the reader's convenience and to facilitate
the reading of the text. 
 
\section{Horizons of de Sitter black hole}
In the subsequent section we will derive the Hawking radiation
Schwarzschild-de Sitter black hole via the surface gravity
$\kappa$ taken at the horizon $r_c$, i.e., $\kappa(r_c)$.
Therefore it makes sense to dwell a little bit on the two
horizons existing in the Schwarzschild-de Sitter case (the elements of
the Schwarzschild- de Sitter metric are given in appendix B).
The starting point here is the horizon condition given by \cite{carroll}:
\begin{equation}   \label{eq:sc}
g^{rr}(r_c)=0
\end{equation}
This condition in the Schwarzschild-de Sitter metric (see appendix B)
is
\begin{equation}   \label{eq:horguide}
1-\frac{2r_s}{r_c}-\frac{1}{3}\frac{r_c^2}{r_\varLambda^2}=0
\end{equation}
where 
\begin{equation}
r_\varLambda=\frac{1}{m_\varLambda}\equiv\frac{1}{\sqrt{\varLambda}}, \,\,\, 
r_s \equiv GM 
\end{equation}
Equation (\ref{eq:horguide}) can be transformed into a third order polynomial
equation, namely
\begin{equation} \label{3rd}
r_c^3-(3r_\varLambda^2)r_c+6r_sr_\varLambda^2=0
\end{equation}
In appendix A we have sketched the solution of a third order
polynomial using an auxiliary angle $\phi$. In the case of equation
(\ref{3rd}) the relevant quantities $p$, $q$, $D$ and $R$ read 
\begin{equation}   \label{eq:1000}
p=-3r_\varLambda^2,\quad q=6r_sr_\varLambda^2, \quad
D=-r_\varLambda^6+9r_s^2r_\varLambda^4< 0, \quad
R=r_\varLambda
\end{equation}
Hence, we can deduce that the polynomial under consideration 
corresponds to the case i) described in the appendix A.
The auxiliary angle can be now defined as
\begin{equation}
\cos\phi=\frac{6r_sr_\varLambda^2}{2r_\varLambda^3}=3\left(\frac{r_s}{r_\varLambda}\right)
\end{equation}
and the solutions are parametrized with the help of trigonometric functions and their inverses
in the following form
\begin{eqnarray}   \label{eq:900}
r_1&=&-2r_\varLambda \cos\left(\frac{1}{3}\cos^{-1}\left(3\frac{r_s}{r_\varLambda}\right)\right) \nonumber \\
r_2&=&-2r_\varLambda \cos\left(\frac{1}{3}\left(\cos^{-1}\left(3\frac{r_s}{r_\varLambda}\right)+2\pi\right)\right)
\nonumber \\
r_3&=&-2r_\varLambda \cos\left(\frac{1}{3}\left(\cos^{-1}\left(3\frac{r_s}{r_\varLambda}\right)+4\pi\right)\right)
\end{eqnarray}
Several conclusions, regarding these solution will be later of importance.
We start to note
that the solutions forbid any result for which $3\frac{r_s}{r_\varLambda}>1$. As a consequence, 
the maximum value of $r_s$ is given by
$3\frac{r_s^{\rm max}}{r_\varLambda}=1$
or, equivalently by
\begin{equation}   \label{eq:rsmax}
r_{s}^{max}=\frac{1}{3}r_\varLambda
\end{equation}
We then obtain a maximum mass , i.e. $M \le M_{\rm max}$ which ensures the
existence of horizons. 
With $G=\frac{1}{m_{\rm pl}^2}$ we can write it as
\begin{equation}   \label{eq:990M}
M_{max}=\frac{1}{3}\frac{m_{\rm pl}^2}{m_\varLambda}
\end{equation}
The next issue of concern is the existence of a maximum horizon i.e. the 
{\it the largest value} $r_3$ can assume while varying the mass $M$. Calculating
$r_i(M_{\rm max})$ gives
\begin{equation}   \label{eq:this}
r_1(M_{\rm max})=-2r_\varLambda, \quad r_2(M_{\rm max})=r_3(M_{\rm max})=r_c^{\rm max}=r_\varLambda 
\end{equation}
From the above we conclude that $r_1$ is an unphysical (indeed, it is always negative), 
and henceforth we keep only $r_2$ and $r_3$ as the relevant physical horizons. 
Note that this is not the maximal horizon in the absolute sense as $r_2$ can   
take values larger than $r_c^{\rm max}$. Indeed, we will see below that
for $M \ll M_{\rm max}$, $r_2$ tends to $\sqrt{3}r_\varLambda$.
However, in the context of the Generalized Uncertainty Principle, discussed in section IV,
it will turn out that $r_c^{\rm max}/\sqrt{10}$ is the largest horizon for a
Schwarzschild- de Sitter black hole with a well defined temperature $T$. Yet another way to confirm the above results is by first 
solving for the mass M from the horizon condition (\ref{eq:horguide}) which gives
\begin{equation}   \label{eq:w2}
M(r_c)=\frac{r_c}{2}m_{\rm pl}^2-\frac{1}{6}\frac{r_c^3}{r_\varLambda^2}m_{\rm pl}^2
\end{equation}
and then looking for a local maximum according to $\partial M/\partial r_c =0$. Solving this equation for the variable $r_c$ results in $r_c^{\rm max}$ as before.
Replacing the previous result in (\ref{eq:w2}), we obtain also as before the maximum mass
defined in equation (\ref{eq:990M}).
Here we can see once again that $M_{\rm max}$ is associated with the maximal horizon $r_\varLambda$.\\
Until now we have approximated the
exact solutions (\ref{eq:900}) 
for the extreme values of the mass, i.e., when $M$ approaches its maximum value $M_{\rm max}$.
However, it will be equally important 
to approximate these solutions for intermediate values of the mass. 
We can first re-write our solution as 
\begin{equation}   \label{eq:r2bs}
r_2\approx -2r_\varLambda \cos\left(\frac{5\pi}{6}-\frac{r_s}{r_\varLambda}
-\frac{3}{2}\left(\frac{r_s}{r_\varLambda}\right)^3\right)
\end{equation}
which can be used to approximate it up to the first order correction 
i.e. $r_2\approx \sqrt{3}r_\varLambda-r_s$. 
Note that this horizon remains non-zero even if we put $M \to 0$ which makes it doubtful that
it is a black hole horizon with a proper temperature.
There is another peculiarity associated with this horizon 
as it decreases with increasing mass. 
It is therefore possible that $r_2$ and $r_3$ 
meet at a certain value of the mass. As mentioned before, they
do that at $M=M_{\rm max}$. We can repeat a similar procedure for $r_3$ obtaining
\begin{equation}   \label{eq:100000}
r_3\approx 2r_s\left(1+\frac{4}{3}\frac{r_s^3}{r_\varLambda^3}\right)
\end{equation}
The correction term proportional to $\frac{r_s^3}{r_\varLambda^3}$ 
is very small in almost the whole range of the masses, 
except in the case when the mass tends to the maximum value given in (\ref{eq:990M}) (for which
the approximated version above is not valid). 
It is often convenient to parametrize 
$r_s=\omega r_\varLambda$ where $\omega$ can take any value between 0 and $\frac{1}{3}$ 
in agreement with (\ref{eq:rsmax}). Then equation (\ref{eq:100000}) reads 
\begin{equation}   \label{eq:1000002}
r_3\approx 2\omega r_\varLambda+\frac{8}{3}\omega^3r_\varLambda
\end{equation}
Even if $\omega \to\omega_{\rm max}=1/3$, the correction term is only of the order $10^{-1}$.

\section{Classical Hawking radiation of de-Sitter black hole}
The idea of this section is the study of the temperature 
as a function of mass for a black-hole in the Schwarzschild-de Sitter metric.
We derive this relation by first calculating the surface gravity $\kappa$ at the horizon $r_c$ and relating 
it to the temperature $T$ of the black hole by $T=\frac{\kappa}{2\pi}$ 
(see \cite{carroll} and \cite{Massadough2}).
The word `classical' refers here exactly to this procedure and we use it to distinguish it
from the results obtained via the Generalized Uncertainty Relation (GUP) in section 4. 
Following the arguments of \cite{carroll} the surface gravity of a black-hole is defined as $\kappa=Va$ where all quantities are evaluated at the horizon $r_c$. 
Here $a$ is the invariant scalar acceleration and $V$ is the red-shift factor, 
which for static observers is equal to the proportionality factor between the timelike Killing 
vector and the four-velocity \cite{carroll} $K^\mu=V(x)U^\mu$. 
Above, $K^\mu$ is the time-like Killing vector and $U^\mu$ is the four-velocity. 
For the Schwarzschild-de Sitter metric we obtain explicitly
\begin{equation}
K^\mu=(1,0,0,0),\;\;\; U^\mu=\left(\left(1-\frac{2r_s}{r}-\frac{1}{3}
\frac{r^2}{r_\varLambda^2}\right)^{-1/2},0,0,0\right)
\end{equation}
and therefore the redshift factor is given by
\begin{equation}   \label{eq:red-shiftfactor}
V=\sqrt{1-\frac{2r_s}{r}-\frac{1}{3}\frac{r^2}{r_\varLambda^2}}
\end{equation}
For the complete evaluation of the surface gravity 
we need the scalar four-acceleration $a$ which we have
derived explicitly in the appendix B.
In the Schwarzschild-de Sitter metric we obtain
\begin{equation}   \label{eq:masifaa}
a=\displaystyle{\frac{\frac{r_s}{r^2}-\frac{1}{3}\frac{r}{r_\varLambda^2}}{\sqrt{1-\frac{2r_s}{r}-\frac{1}{3}
\frac{r^2}{r_\varLambda^2}}}}
\end{equation}
Hence the surface gravity takes the following simple expression
\begin{equation}   \label{eq:happa}
\kappa(r_c)=\left|\frac{r_s}{r_c^2}-\frac{1}{3}\frac{r_c}{r_\varLambda^2}\right|
\end{equation}
We will comment about the absolute value in this expression in the next sub-section. 
The expression (\ref{eq:happa}) can be further simplified
such that $\kappa$ is a function of the horizons alone \cite{Xiang}. 
This can be achieved by the replacing equation (\ref{eq:w2}) into (\ref{eq:happa}) .
The final formula is then
\begin{equation}   \label{eq:thistemp990}
\kappa(r_c)=\left\vert\frac{1}{2r_c^2}\left(r_c-\frac{1}{3}\frac{r_c^3}{r_\varLambda^2}\right)
-\frac{1}{3}\frac{r_c}{r_\varLambda^2}\right\vert
=\left\vert\frac{1}{2r_c}-\frac{1}{2}\frac{r_c}{r_\varLambda^2}\right\vert
\end{equation}
Viewing $\kappa$ as a function of the horizon or alternatively as a function of mass (see (\ref{eq:happa}))
we note that $\kappa(r_c^{\rm max})=\kappa(M_{\rm max})=0$ 
which obviously implies $T(M_{\rm max})=0$ by the virtue of $T=\kappa/2\pi$.
By the same identification between surface gravity and temperature the full
$T-M$ relation can be spelled out as
\begin{equation} \label{extra2}
T(M)=\frac{\kappa(r_3)}{2\pi}=\frac{1}{2\pi}\left(\frac{\cos\left(\frac{1}{3}
\left(\cos^{-1}\left(3\frac{r_s}{r_\varLambda}\right)+4\pi\right)\right)}{r_\varLambda}-
\frac{1}{4r_\varLambda \cos\left(\frac{1}{3}
\left(\cos^{-1}\left(3\frac{r_s}{r_\varLambda}\right)+4\pi\right)\right)}\right)
\end{equation}
In the case of $r_c=r_3$ the absolute value, which appears in (\ref{eq:thistemp990}),  
is not necessary as $T(M_{\rm max}) = 0$. We will discuss the case $\kappa(r_2)$ in a suitable approximation in the next sub-section. However,
already here we note that this case is physically not without inconsistencies.
As $M$ increases, the temperature will decrease in this case, but the horizon will become smaller
and therefore also the entropy. 

\subsection{First order corrections of $\Lambda$ in Hawking radiation}   
In the approximate version $r_3 \approx 2r_s$ the formula (\ref{eq:thistemp990}) simplifies considerably
and gives us the first order correction to the standard Hawking expression:
\begin{equation}
\kappa(r_3)=\frac{m_{\rm pl}^2}{4M}-\frac{m_\varLambda^2}{m_{\rm pl}^2}M
\end{equation}
The $T-M$ relation reads
\begin{equation}   \label{eq:why?}
T(M)=\frac{m_{pl}^2}{8\pi M}-\frac{1}{2\pi}\frac{m_\varLambda^2}{m_{\rm pl}^2}M
\end{equation}
valid for every $M$, except as $M\to M_{\rm max}$ given in (\ref{eq:990M}).
In fact, the case of the the maximum value is included already in the equation $\kappa(r_c^{\rm max})=\kappa(M_{\rm max})=0$. Note that in (\ref{eq:why?}) we have not used the absolute value.
The reason is that for every mass the temperature defined 
in this way (for the horizon $r_3$) is positive. 
This in turn is a consequence of the equivalence principle \cite{Massadough2}, 
because a local inertial observer in comparison with a static one 
perceives a positive scalar acceleration calculated from (\ref{eq:masifaa}). 
The opposite happens for the horizon $r_2$. 
In that case, for every value taken by $r_2$, except that obtained in (\ref{eq:this}), 
the surface gravity given in (\ref{eq:thistemp990})
would be negative without taking the absolute value. 
The reason for this behavior is again due to the equivalence principle 
because in this case the local inertial observer is moving 
while $r$ increases. Therefore the static observer 
has a negative scalar acceleration. 
For example the maximum value of $r_2$ 
is $\sqrt{3}r_\varLambda$, in which case the result (\ref{eq:r2bs}) replaced in (\ref{eq:thistemp990}), 
gives
$\kappa=\left\vert-\frac{1}{\sqrt{3}}m_\varLambda\right\vert=\frac{1}{\sqrt{3}}m_\varLambda$.
On the other hand, for intermediate values of the mass, 
replacing $r_2\approx\sqrt{3}r_\Lambda-r_s$ in (\ref{eq:thistemp990}) leads to
\begin{equation} \label{kappaxxx2}
\kappa(r_2)\approx \left\vert \frac{-2r_\varLambda^2+2\sqrt{3}r_\varLambda r_s
-r_s^2}{2(\sqrt{3}r_\varLambda-r_s)r_\varLambda^2}\right\vert
\approx \frac{r_\varLambda-\sqrt{3}r_s}{r_\varLambda(\sqrt{3}r_\varLambda-r_s)}
\end{equation}
The temperatures associated with the above results is, respectively
\begin{equation}   \label{eq:anoT2}
T(M<<M_{\rm max})\approx\frac{1}{2\pi\sqrt{3}}m_\varLambda
\end{equation}
and
\begin{equation}   \label{eq:anoT}
T(M)\approx \frac{1}{2\pi}\left(\frac{r_\varLambda-\sqrt{3}r_s}{r_\varLambda(\sqrt{3}r_\varLambda-r_s)}\right)
\end{equation}
Apparently the behavior of the temperature function given in (\ref{eq:anoT}) is
correct. In fact as $M$ increases, the temperature decreases , i.e., 
$\frac{\partial T}{\partial M}<0$ as it should be 
in view of the fact that the heat capacity of the black-hole is negative. 
The same is accomplished by (\ref{eq:why?}). 
Recall that we have insisted here on the absolute value because 
the acceleration of the static observer with respect to a local one is negative.
Without the absolute value in the expression (\ref{kappaxxx2})
we would have negative temperatures and a positive slope $\frac{\partial T}{\partial M}>0$ for the function $T(M)$. 
However, a different inconsistency appears now in the entropy behavior.
An increase of mass in $r_2\approx\sqrt{3}r_\Lambda-r_s$ 
implies a decrease of the horizon and as a consequence of that 
a decrease in the standard entropy value \cite{carroll}. 
With these arguments in mind it is reasonable to discard the temperature function due to 
the horizon $r_2$.

\subsection{Consequences}
In the next section we will elaborate on the problem of Hawking radiation
of Schwarzschild-de Sitter black hole from the point of view of
the Generalized Uncertainty Principle (GUP). It therefore makes sense to
collect here the important results we obtained the the preceding sections. 
The results (\ref{extra2}) and (\ref{eq:why?}) associated with $\kappa(r_3)$
represent  the correct physical behavior. 
The heat capacity of the black hole is negative and the horizon 
and the entropy increases with the mass. The latter aspect is missing for
$\kappa(r_2)$.
The Schwarzschild radius $2r_s$ has a maximum allowed value $2r_\varLambda/3$ (see eq.
(\ref{eq:rsmax})) corresponding to a 
maximum mass given in (\ref{eq:990M}) which gives a maximum allowed horizon $r_c^{\rm max}=
r_\varLambda$ in eq. (\ref{eq:100000}). At the maximum mass (or horizon) the Hawking temperature
becomes zero. This we can interpret as a minimum temperature in this case.
We mention this explicitly since $T_{\rm min}$ will come out non-zero using the
GUP approach below.

\section{Hawking radiation via the Generalized Uncertainty Principle (GUP)}   
In this section we consider the Hawking radiation via GUP 
developed for the case $\varLambda=0$ in \cite{adler1} and \cite{adler2} (see also 
\cite{GUPimportant3} and \cite{GUP3important}).
The main result is the deformation of the $T-M$ dispersion relation
close to the Planck length $l_{\rm pl}$ which turns out to be now the minimum
possible horizon. Another way of expressing this result is to say that the black hole mass
$M$ has a remnant of the order of Planck mass $m_{\rm pl}$ (this defines also the
minimum possible mass). With the inclusion of $\varLambda$ we have seen in the preceding section 
that there exist a maximum horizon and maximum mass. The simple question which we pose
here in connection with $\varLambda$ is whether the $T-M$ relation gets modified also
close to $r_c^{\rm max}$ (or, which is equivalent, close to $M_{\rm max}$).
In this context it is worth noting that gravity with $\varLambda$ displays often
a duality. Where the Newtonian constant $G$ sets a minimum (maximum) allowed value,
the cosmological constant $\varLambda$ restricts the range of a parameter by setting
a maximum (minimum). An example is the range of validity of the Newtonian limit  \cite{we}.
Here the distance $r$ is limited by 
\begin{equation} \label{Nlimitrange}
2r_s \approx r_3 \ll r \ll r_2 \approx \sqrt{3}r_\varLambda 
\end{equation}
for the intermediate mass range $M$.
Another example of such duality is encountered in the motion of a test particle
in the Schwarzschild-de Sitter metric \cite{we2}. The equation of motion can be brought into
the form containing an effective potential $U_{eff}$ which depends parametrically on 
$r_s$, $r_\varLambda$ and the angular momentum per mass $r_l$. The effective potential
has generically three local extrema: a maximum close to $2r_s$, a minimum in which the planets move and,
due to $\varLambda$, a second maximum. To avoid that the first local maximum and the local minimum
coincide to form a saddle point, one has to respect the inequality 
\begin{equation} \label{lmin}
r_l > r_l^{\rm min}=2\sqrt{3r_s}
\end{equation}
On the other hand, if we insist that the local minimum and maximum do not
degenerate to a saddle point, we have to satisfy \cite{we2}
\begin{equation} \label{lmax}
r_l < r_l^{\rm max}=\left(\frac{3}{4}\right)^{1/3}(r_s^2 r_\varLambda)^{1/3}
\end{equation}
It is not unreasonable to expect that the Hawking radiation of Schwarzschild-de Sitter black hole displays
similar duality features. 
\subsection{GUP with $\varLambda=0$}
It makes sense to have first a brief glimpse at the case $\Lambda=0$.
The Generalized Uncertainty Principle (GUP) \cite{adler1, GUP2, GUP3} and 
the discussion of Hawking radiation within its framework \cite{adler2, GUPBH1, GUPBH2} has gained 
some popularity in the last few years. Therefore while discussing the case $\varLambda=0$
we will only give the
main steps which are of importance in generalizing it to $\varLambda \neq 0$.
  
The steps involved in deriving the uncertainty relation with 
gravity are \cite{adler1, adler2} (i) $\Delta x_{\rm grav} \sim
(\vert \vec{F}_{\rm grav}\vert /m)L^2$, where $L$ is the typical length/time scale and here 
$\vert \vec{F}_{\rm grav}/m\vert=r_s/r^2$, (ii) $E$ being the photon's energy is the source of gravity felt
by the probed particle; $E=p \sim \Delta p$ which is the uncertainty in momentum of the latter, 
(iii) $r \sim L$ taken together with the previous 
steps gives now $\Delta x_{\rm grav} \sim G_N \Delta p$ which is to be added to the standard uncertainty
relation resulting in
\begin{equation} \label{standard}
\Delta x \gtrsim \frac{1}{2\Delta p} + \frac{\Delta p}{2 m_{\rm pl}^2}
\end{equation}
It is evident that the method of obtaining (\ref{standard}) is heuristic (we follow here
\cite{adler1}). This is, however, not a drawback as the same result is obtained within
string theory (see the papers by G. Veneziano in \cite{GUP2}) 
and non-commutative geometry (see the papers by M. Maggiore in \cite{GUP2}). This
shows that the heuristic line of arguments is indeed valid and has the advantage of being
also model independent.
 
Applying (\ref{standard}) to black hole evaporation \cite{adler2} 
consists essentially in identifying $\Delta x$ with the Schwarzschild
radius \cite{Ruffini} 
($2r_s$ in our notation) as well as $\Delta p \sim p =E$ with the temperature up to a factor. 
This turns out to be the surface gravity $T_*=\kappa$ such that $T=T_*/2\pi$.
The result is a quadratic equation in $T_*$
\begin{equation} \label{quadratic}
T_*^2 -4MT_*+\frac{m_{\rm pl}^2}{2}=0
\end{equation}
from which it follows that
\begin{equation} \label{solution}
T_*(M)=2M\left(1-\sqrt{1-\frac{m_{\rm pl}^2}{4M^2}}\right)\to \frac{m_{\rm pl}^2}{4M}
\end{equation}
where we have chosen already a solution with the correct limit at large $M$ as indicated in the
above equation.
Two conclusions are in order. First, the temperature is well-defined only if
\begin{equation} \label{ram1}
M > M_{\rm min}=M_{\rm remnant}=\frac{m_{\rm pl}}{2}
\end{equation}
which defines the minimum mass and the black hole remnant. 
Secondly, the existence of a minimum mass sets a scale for the maximally possible temperature $T_{\rm max}$
via (\ref{solution})  and $T=T_*/2\pi$. It reads
\begin{equation}   \label{eq:maxumus}
T_{\rm max}=\frac{M_{\rm min}}{\pi}=\frac{m_{\rm pl}}{2 \pi}
\end{equation}
Yet another interpretation of the above results refers to the length scales involved. 
The existence of a black hole remnant is equivalent 
to say that the Schwarzschild horizon can not be smaller than the Planck scale  
$l_{\rm pl}=1/m_{\rm pl}$ \cite{adler1, adler2, GUP3important, A.E.}, i.e., $2r_s^{\rm min}=l_{\rm pl}$. 
This can be easily verified by re-writing (\ref{solution}) as
\begin{equation}   \label{eq:minus233}
T_*=2\frac{r_s}{l_{\rm pl}^2}\left(1-\sqrt{1-\frac{1}{4}\left(\frac{l_{\rm pl}}{r_s}\right)^2}\right)
\end{equation}
which is well defined for $r_s > r_s^{\rm min}$. Note that this minimum length scale is exactly what
one would expect from quantum gravity. However, we should not forget
that any estimation deduced from an uncertainty relation like (\ref{quadratic}) remains
an order of magnitude estimate having at the same time the advantage of being model independent. 
Choosing the right branch among the two
solutions of the quadratic equation (\ref{quadratic}) has, as mentioned above, to do with
the right limit for large masses which is known by the Hawking formula. However, even without
knowing this limit explicitly, we could discriminate the physical solution from the non-physical one
by using arguments based on the negative heat capacity of the Schwarzschild black-hole, i.e., 
insisting on $\frac{\partial T}{\partial M}<0$ The latter is a consequence of $\frac{\partial S}{\partial T}<0$. 
This together with $\frac{\partial S}{\partial M}>0$ allows us to conclude that $\frac{\partial T}{\partial M}<0$ 
on very general grounds. This expectation is satisfied only if we choose the right physical
solution of (\ref{quadratic}). Indeed, we obtain then
\begin{equation}   \label{eq:slopeMass}
\frac{dT_*}{dM}=\frac{2}{\sqrt{1-\frac{1}{4}\left(\frac{m_{\rm pl}}{M}\right)^2}}
\left(\sqrt{1-\frac{1}{4}\left(\frac{m_{\rm pl}}{M}\right)^2}-1\right)<0
\end{equation}
since we have $\sqrt{1-\frac{1}{4}\left(\frac{m_{\rm pl}}{M}\right)^2}\leqslant1$. 
As we have already seen in subsection A such general restriction are often
not unimportant to exclude a possible solution.
\subsection{GUP with $\varLambda \neq 0$}
The Generalized Uncertainty Principle with $\varLambda=0$ bears interesting results in agreement
with expectation from quantum gravity. The dispersion relation $T(M)$ gets modified near the Planck radius
$2r_s^{\rm min}=l_{\rm pl}$ as
compared to the standard Hawking result. We can paraphrase this also
by stating that there exists a minimum mass $M_{\rm min}$ which corresponds to a maximum temperature
$T_{\rm max}$. Motivated by the duality encountered in gravity theory with $\varLambda$ (see
eqs. (\ref{lmin}) and (\ref{lmax})), we can speculate that the Generalized Uncertainty Principle
with $\varLambda  \neq 0$ will give us a dual relation where 
$T(M)$, as compared to (\ref{eq:why?}), is modified close
to $M \sim M_{\rm max}$ (eq. \ref{eq:990M}) (or, which is equivalent, close to $r_s^{\rm max}$ from eq. 
(\ref{eq:rsmax})). This should give us a $T_{\rm min} \neq 0$ given by the scales of $\varLambda$.
Anticipating our results, we mention already here that this is indeed the case and we obtain
$T_{\rm min} \sim m_\varLambda$.

We have seen that the Generalized Uncertainty Principle can be obtained easily from the
gravitational force. 
One can repeat this heuristic approach with GUP including $\varLambda \neq 0$. Since this 
uncertainty relation is new, we will check the results emerging from it by comparing it with
(i) standard results for $T(M)$ and (ii) independent results in the context of black body radiation
in section five. Both these checks will show that the new GUP relation is consistent. To repeat the steps leading to GUP from the previous section we need
the gravitational potential $\Phi$ for a spherically symmetric mass distribution with $\varLambda$
\cite{we}
\begin{equation} \label{potential}
\Phi=-\frac{r_s}{r}-\frac{1}{6}\frac{r^2}{r_{\varLambda}^2}
\end{equation} 
Then following the arguments from the last sub-section the gravitational force per mass
attributed to $\varLambda$ is $\frac{\vert\vec F_\varLambda\vert}{m}=\frac{1}{3}\varLambda L$ where $L$ is again a typical length scale in the problem under consideration. The corresponding displacement is $\Delta x_\varLambda\thicksim \frac{1}{3}m_\varLambda^2L^3$. 
We use now the additional  assumption $L\thicksim\frac{1}{\Delta p}$ \cite{makh}.
This assumption is equivalent to say that the precision of the momentum is
inversely proportional to the typical length scale and can be found e.g. in
\cite{LL2, messiah} in connection with wave packets. It is analog to similar 
assumptions like $\Delta t \thicksim E^{-1}$ in the context of
estimating the pion mass in Yukawa's theory \cite{gasiorowicz} or $\Delta x \thicksim p^{-1}$
in case we want to estimate the precision of the position \cite{LL4}. Therefore we can write $\Delta x_\varLambda\thicksim \frac{1}{3}\frac{m_\varLambda^2}{\Delta p^3}$ such that the 
proposed relation for GUP with the inclusion of the cosmological constant is
\begin{equation} \label{eq:findel} 
\Delta x \gtrsim \frac{1}{2\Delta p}+\frac{\Delta p}{2m_{\rm pl}^2}-\Delta x_\varLambda,
\quad
\Delta x \gtrsim \frac{1}{2\Delta p}+\frac{\Delta p}{2m_{\rm pl}^2}
-\frac{\gamma}{3}\frac{m_\varLambda^2}{\Delta p^3}
\end{equation}
where we have taken into account the relative sign difference between the cosmological constant 
contribution and the standard Newtonian part \cite{we}. 
We also include a factor $\gamma \thicksim {\cal O}(1)$ which accounts for the fact that
we are dealing with orders of magnitudes estimates. In comparing the results
with (\ref{eq:why?}) for masses smaller than $M_{\rm max}$, $\gamma$ should come out of the order of $1$.
If this is not the case, something would be wrong with the uncertainty relation
(\ref{eq:findel}).
As in the previous sub-section in the context of Hawking radiation 
the uncertainty in position is associated with the event horizon. 
In the case of Schwarzschild-de Sitter metric, we should, in principle, 
take the full expression (\ref{eq:100000}). It will turn out, however, that it is sufficient
to use the approximation $2r_s$.
Then the Generalized Uncertainty applied to black hole evaporation 
gives an equation which generalizes (\ref{quadratic})
\begin{equation}   \label{eq:thriumph}
\frac{2M}{m_{\rm pl}^2} = \frac{1}{2 T_*}+\frac{T_*}{2m_{\rm pl}^2}-\frac{\gamma}{3}\frac{m_\varLambda^2}{T_*^3}
\end{equation}
It is worth noting that
for high temperatures, the results of the previous sub-section for $\varLambda=0$ are recovered from
(\ref{eq:thriumph}).
Therefore, $T_{\rm max}$ in conjunction with $M_{\rm min}$ also follows from the above equation.  
For small temperatures (\ref{eq:thriumph}) can be approximated to
\begin{equation}   \label{eq:GUP9902}
\frac{2M}{m_{\rm pl}^2}\approx \frac{1}{2T_*}-\frac{\gamma}{3}\frac{m_\varLambda^2}{T_*^3}
\end{equation}
which amounts to solve a third order  polynomial
of the form
\begin{equation} \label{eq:toeGUP990}
T_*^3-\left(\frac{m_{\rm pl}^2}{4M}\right)T_*^2+\frac{\gamma}{6}\frac{m_\varLambda^2 m_{\rm pl}^2}{M}\ =0
\end{equation}
To solve this equation we refer to appendix A. In connection with (\ref{eq:toeGUP990}) the following auxiliary constants are needed $
r=-\frac{m_{\rm pl}^2}{4M}$, $s=0$, $t=\frac{\gamma}{6}\frac{m_\varLambda^2 m_{\rm pl}^2}{M}$ to obtain
the reduced form of (\ref{eq:toeGUP990}) 
which is reached by the shift
\begin{equation}   \label{eq:yvar}
y=T_*+\frac{r}{3}=T_*-\frac{m_{\rm pl}^2}{12M}
\end{equation}
where $y$ is the solution of the reduced third-order equation given in appendix A. 
The coefficients of the reduced equation can be calculated explicitly. They are $p=-\frac{m_{pl}^4}{48M^2}$ and
\begin{equation}   \label{eq:qG2}
q=\frac{m_{\rm pl}^4}{M}\left(-\frac{1}{864}\frac{m_{\rm pl}^2}{M^2}
+\frac{\gamma}{6}\frac{m_\varLambda^2}{m_{\rm pl}^2}\right)
\end{equation}
The parametric solution depends on the sign of $q$ which depends only on the variable $M$.
We denote the branch point by $M_{q=0}$ which can be found by setting $q=0$.
We find
\begin{equation}   \label{eq:mq990}
M_{q=0}=\frac{1}{12\sqrt{\gamma}}\frac{m_{\rm pl}^2}{m_\varLambda}
\end{equation}
such that $M<M_{q=0}$ for $q<0$ and $M>M_{q=0}$ for $q>0$. The existence of real solution i.e. $T_*(M)$ or $y$ 
depends  crucially 
on $D$ in appendix A. In the case under consideration it reads
\begin{equation}
D=\frac{1}{4}\frac{m_{\rm pl}^6 m_\varLambda^2}{M^2}
\left(\frac{\gamma^2}{36}\left(\frac{m_\varLambda^2}{m_{\rm pl}^2}\right)
-\frac{\gamma}{3(864)}\left(\frac{m_{\rm pl}^2}{M^2}\right)\right)
\end{equation}
It can be demonstrated that for $D>0$ there are no physical solutions 
of the associated third-order equation and 
only $D<0$ is of interest for us. 
A limit on the value of $M$ is set by putting $D=0$. We find from $D=0$,
$M_{\rm max}^*$, 
\begin{equation}   \label{eq:maxm990}
M^*_{\rm max}=\frac{1}{6\sqrt{2\gamma}}\frac{m_{\rm pl}^2}{m_\varLambda}
\end{equation}
such that $M< M_{\rm max }^*$ if $D <0$.
Later in text we will find $\gamma=5/9$ by comparing the GUP solution $T(M)$ 
to the one found in section III. 
In other words, we have also $D(M> M_{\rm max}^*) > 0$. The real solution in the case $p <0$ and $D > 0$ 
(see case ii) in appendix A) is $y_1=-2R\cosh \frac{\phi}{3}$ which is positive definite if $R <0$.
The latter implies $q <0$ and from this we conclude that $M < M_{q=0}$. However, as we will show below,
we have $M_{\rm max}^* > M_{q=0}$ which is in contradiction to $D(M> M_{\rm max}^*) > 0$.
Opting for $q > 0$ (i.e. $R >0$) the solution is $T_{1*}=y_1 + R=R(1-2\cosh \phi/3)$ which is always
negative since the smallest value of $\cosh x $ is $1$. 
A remark about the three different mass scales is in order.
We have
\begin{equation} \label{extra3}
M_{\rm max} > M_{\rm max}^* > M_{q=0}
\end{equation}
where $M_{\rm max}$ is the value found in (\ref{eq:990M}) in connection with $r_s^{\rm max}$.
Nevertheless all these values are of the same order of magnitude as
\begin{equation}   \label{eq:mq02}
M_{q=0}=\frac{M^*_{max}}{\sqrt{2}} \approx \frac{M_{\rm max}}{2\sqrt{2}}
\end{equation}
The correction to $2r_s$ at $M=M_{\rm max}^*$ given in (\ref{eq:100000}) 
as $4r_s^3/3r_\varLambda^3$ is suppressed by
one order of magnitude as compared to $1$. This justifies the use of $2r_s$ 
as an approximation in equation (\ref{eq:GUP9902}). 
It will be convenient from now on to parametrize the mass $M$ by a parameter $\zeta$ defined by
\begin{equation}   \label{eq:pareq}
M=\frac{M^*_{\rm max}}{\zeta}
\end{equation}
where $\zeta=1$ corresponds to $M^*_{\rm max}$.
The branch point corresponding to $q=0$ can be now characterized 
by $\zeta>\sqrt{2}$ for $q<0$ and $1<\zeta <\sqrt{2}$ for $q>0$.
\subsubsection{The branch $q> 0$}
The parameter R given in appendix A, 
which depends on the sign of $q$ is simply
\begin{equation}   \label{eq:RGUP}
R=\frac{1}{12}\frac{m_{\rm pl}^2}{M}
\end{equation}
Obviously, case i) from appendix A applies in this case.
Hence the auxiliary angle, as $D<0$ and $p<0$, can be calculated as
\begin{equation}   \label{eq:aagup990}
\cos\phi=-1+144\gamma\frac{M^2m_\varLambda^2}{m_{\rm pl}^4}
\end{equation}
The zeros of the reduced third-order equation in terms of the parameter $\zeta$ in (\ref{eq:pareq})
can be easily found to be
\begin{equation}
y_1=-\sqrt{2\gamma}m_\varLambda\zeta \cos\left(\frac{1}{3}\cos^{-1}\left(-1+\frac{2}{\zeta^2}\right)\right)
\end{equation}
\begin{equation}
y_2=-\sqrt{2\gamma}m_\varLambda\zeta \cos\left(\frac{1}{3}
\left(\cos^{-1}\left(-1+\frac{2}{\zeta^2}\right)+2\pi\right)\right)
\end{equation}
\begin{equation}
y_3=-\sqrt{2\gamma}m_\varLambda\zeta \cos\left(\frac{1}{3}\left(\cos^{-1}
\left(-1+\frac{2}{\zeta^2}\right)+4\pi\right)\right)
\end{equation}
From equations (\ref{eq:yvar}) and (\ref{eq:pareq})
it is possible to find the explicit solutions for the surface gravity $T_*$:
\begin{equation}   \label{eq:maly}
T_{1*}(\zeta)=-\sqrt{2\gamma}m_\varLambda\zeta\left(\cos\left(\frac{1}{3}\cos^{-1}
\left(-1+\frac{2}{\zeta^2}\right)\right)-\frac{1}{2}\right)
\end{equation}
\begin{equation}   \label{eq:t2star}
T_{2*}(\zeta)=-\sqrt{2\gamma}m_\varLambda\zeta\left(\cos
\left(\frac{1}{3}\left(\cos^{-1}\left(-1+\frac{2}{\zeta^2}\right)+2\pi\right)\right)-\frac{1}{2}\right)
\end{equation}
\begin{equation}   \label{eq:t3}
T_{3*}(\zeta)=-\sqrt{2\gamma}m_\varLambda\zeta\left(\cos
\left(\frac{1}{3}\left(\cos^{-1}\left(-1+\frac{2}{\zeta^2}\right)+4\pi\right)\right)-\frac{1}{2}\right)
\end{equation}
It remains to discuss which of the above solutions is physical (bearing in mind that the real
temperature $T$ is $T=T_* /2 \pi$). It is easy to show that $T_1=T_{1*}/2\pi$ is negative.
The right choice between the solutions can be done by the requirement that the
deformed $T(M)$ relation must smoothly match the classical result for moderate masses i.e.
equation (\ref{eq:why?}) (which in turn for even smaller masses goes over to the
standard Hawking formula). This is impossible if the deformed solution increases with mass as
(\ref{eq:why?}) has negative heat capacity. 
One can show that $\partial T_3/\partial M$ is always positive and therefore
can be discarded as a physical solution.
To see that, it suffices to calculate $T_2$ as well as  $T_3$ at two different points.
We start with $\zeta= 1$ ($M=M^*_{\rm max}$) where we consider the surface gravity as a function of $\zeta$. We get
\begin{equation}   \label{eq:again990}
T_{2*}(1)=T_{3*}(1)=T_{\rm min *}=\sqrt{2\gamma}m_\varLambda
\end{equation}
If we can establish that the physical solution is $T_2$,
this would imply the existence of a minimum temperature due to $\varLambda$
in conjunction with $M_{\rm max}^*$ at a horizon approximately $r_\varLambda /3$, namely
\begin{equation}   \label{eq:tmin}
T_{\rm min}=\frac{T_{2*}(1)}{2\pi}
=\frac{T_{\rm min *}}{2\pi}=\frac{\sqrt{\gamma}}{\sqrt{2}\pi}m_\varLambda\approx 0.225\sqrt{\gamma} m_\varLambda
\end{equation}
At $\zeta\to\sqrt{2}_{-}$, where the sub-index '-' implies the limit taken from the left, 
we have $T_{2*}(\sqrt{2}_-)=2.73\sqrt{\gamma}m_\varLambda$ and $T_{3*}=\sqrt{\gamma}m_\varLambda$. 
These results are equivalent to $T_{2}(\sqrt{2}_-)=0.4348\sqrt{\gamma}m_\varLambda\;\;>T_{\rm min}$ 
and $T_{3}(\sqrt{2}_-)=0.159\sqrt{\gamma}m_\varLambda\;\;<T_{\rm min}$. 
Hence $T_3$ a monotonically increasing function with $M$ and therefore $T_2$ is the physical
solution. This establishes $T_{\rm min}$ in (\ref{eq:tmin}) as a genuine minimal value of
temperature.

\subsubsection{The branch $q< 0$}
In this case the auxiliary angle is 
\begin{equation}   \label{eq:aagup2}
\cos\phi=1-144\gamma\frac{M^2m_\varLambda^2}{m_{\rm pl}^4}
\end{equation}
Following the same procedure as above we arrive at
\begin{equation}   \label{eq:t1s}
T_{1*}^{\prime}(\zeta)=\sqrt{2\gamma}m_\varLambda\zeta\left(\cos
\left(\frac{1}{3}\cos^{-1}\left(1-\frac{2}{\zeta^2}\right)\right)+\frac{1}{2}\right)
\end{equation}
\begin{equation}   \label{eq:t2s}
T_{2*}^{\prime}(\zeta)=\sqrt{2\gamma}m_\varLambda\zeta
\left(\cos\left(\frac{1}{3}\left(\cos^{-1}\left(1-\frac{2}{\zeta^2}\right)+2\pi\right)\right)+\frac{1}{2}\right)
\end{equation}
\begin{equation}   \label{eq:t32}
T_{3*}^{\prime}(\zeta)=\sqrt{2\gamma}m_\varLambda\zeta\left(\cos
\left(\frac{1}{3}\left(\cos^{-1}\left(1-\frac{2}{\zeta^2}\right)+4\pi\right)\right)+\frac{1}{2}\right)
\end{equation}
To check the continuity at the branch point 
it is necessary to evaluate these equations at $\zeta\to \sqrt{2}_+$, 
where the sub-index '+' denotes the limit approached from the right. 
We obtain the following equalities $T_{2*}(\zeta\to\sqrt{2}_-)=T_{1*}^{\prime}(\zeta\to\sqrt{2}_+)$ and $T_{3*}(\zeta\to\sqrt{2}_-)=T_{3*}^{\prime}(\zeta\to\sqrt{2}_+)$ which means that on this branch $T_{1*}^{\prime}$ is the physical solution.
After some expansion the $T-M$ relation in the vicinity of $\zeta = \sqrt{2}$
is approximately given by
\begin{equation}   \label{eq:agotr}
T(M) \approx T_{1}^{\prime}\approx\frac{1}{8\pi}\frac{m_{\rm pl}^2}{M}
-\frac{10}{9\pi}\left(\frac{m_\varLambda}{m_{\rm pl}}\right)^2M
\end{equation}
The fact that $T_{1*}^{\prime}$ is the right physical choice can also be established
by expanding the above results for
$\zeta \gg 1$. From (\ref{eq:t1s}), (\ref{eq:t2s}) and (\ref{eq:t32}) it follows $T_{1*}^{\prime}(\zeta \gg 1) \approx \frac{3\sqrt{2\gamma}}{2}m_\varLambda\zeta$ and $T_{2*}=T_{3*} \approx 0$ confirming again that (\ref{eq:t1s}) is the right 
physical choice on this branch 
since replacing the definition of $\zeta$ therein and using $T=T_*/2\pi$ we obtain the
standard Hawking result. i.e., 
\begin{equation} \label{hawking} 
T(M)\approx T_1=\frac{1}{8\pi}\frac{m_{\rm pl}^2}{M}
\end{equation}
Note that our starting point has been the uncertainty relation (\ref{eq:thriumph}) which we 
have already approximated in
(\ref{eq:GUP9902}) for small temperature. Although, it is easy to see that the full uncertainty relation
(\ref{eq:thriumph}) approximated for large temperatures leads to the same results as discussed in
sub-section A (this refers especially to the existence of $T_{\rm max}$ and $M_{\rm min}$),
the Hawking result (\ref{hawking}) is the extreme low mass expansion which follows from the 
approximation (\ref{eq:GUP9902}).  For the intermediate mass range we would expect that we recover
the functional form of equation (\ref{eq:why?}). This will allow us to fix, in principle,  the parameter $\gamma$
and to make a consistency check of the uncertainty relation (\ref{eq:findel}). 
Recall that arguments in connection with uncertainty relations involve orders of magnitude estimates.
Therefore, if we recover form uncertainty relation the functional from of (\ref{eq:why?}) such that
in comparison with (\ref{eq:toeGUP990}) we get $\gamma \sim {\cal O}(1)$, 
then the uncertainty relation (\ref{eq:findel}) is certainly consistent.
We elaborate on these issues in more detail below.

\subsubsection{The matching condition}
We start with 
\begin{equation}   \label{eq:t1app}
T_{*}(M)=\frac{1}{12}\frac{m_{\rm pl}^2}{M}+\frac{1}{6}\frac{m_{\rm pl}^2}{M}\cos
\left(\frac{1}{3}\cos^{-1}\left(1-2\left(\frac{M}{M^*_{\rm max}}\right)^2\right)\right)
\end{equation}
obtained from $T_{1*}^{\prime}$.
For $M<<M^{*}_{\rm max}$, we can expand (\ref{eq:t1app}) in powers of $M/M^{*}_{max}$  and divide $T_*$ by $2\pi$ to arrive at
\begin{equation}   \label{eq:ttriste}
T_{*}(M)\approx \frac{1}{8\pi}\frac{m_{\rm pl}^2}{M}
-\frac{9}{10\pi}\gamma\left(\frac{m_\varLambda}{m_{\rm pl}}\right)^2M
\end{equation}
We see that the functional form is identical to (\ref{eq:why?}).
Demanding that in this mass region the two results be equal
allows us to fix $\gamma=5/9$. Indeed, we see that $\gamma$ is of order $1$ as it should be if the uncertainty relation 
with $\varLambda$ is physically relevant. Fixing the parameter $\gamma$ permits us to write  
\begin{equation}   \label{eq:temin990}
T_{\rm min}=1.05m_{\varLambda} \approx m_\varLambda, \,\,\, M_{\rm max}^* =\frac{1}{2\sqrt{10}}\frac{m_{\rm pl}^2}
{m_\varLambda}
\end{equation}
Recall that we started probing into the Hawking radiation via the Generalized Uncertainty relation
to see if by the inclusion of $\varLambda$ we get a relation
dual to $T_{\rm max}$ and $M_{\rm min}$ which is a result of GUP with $\varLambda=0$ (and also
with $\varLambda \neq 0$ for large temperatures). Equation (\ref{eq:temin990}) is indeed
such a dual result due to $\varLambda$.

\subsubsection{GUP with $\varLambda\neq0$ in the extreme value of the horizon}
In equation (\ref{eq:thriumph}) we used for the horizon the value $2r_s$ with the justification
that $r_3(M_{\rm max}^*) \approx 2r_s(M_{\rm max}^*)$ with correction being roughly $10^{-1}$.
Nevertheless, this does not exclude the possibility that there exists a solution
$T(M)$ if we start in equation (\ref{eq:findel}) with a horizon bigger than
$2r_s(M_{\rm max}^*)$. For the sake of completeness,
we probe into this matter by parametrizing the horizon as $\beta r_\varLambda$ where we are
interested in the parameter $\beta$ around the value $1$. The relevant equation, corresponding to
(\ref{eq:toeGUP990}), is now
\begin{equation}   \label{eq:findel990}
\beta r_\varLambda\approx \frac{1}{2T_*}-\frac{5}{27}\frac{m_\varLambda^2}{T_*^3}
\end{equation}
The third order equation takes the form
\begin{equation}
T_*^3-\left(\frac{m_{\varLambda}}{2\beta}\right)T_*^2+\frac{5}{27\beta}m_{\varLambda}^3=0
\end{equation}
with $r=-\frac{m_{\varLambda}}{2\beta}$ and $t=\frac{5}{27}\frac{m_\varLambda^3}{\beta}$ being the coefficients of the reduced third order polynomial. According to appendix A we have 
\[
y=T_*-\frac{m_{\varLambda}}{6\beta},\quad p=-\frac{m_{\varLambda}^2}{12\beta^2},\quad  q=\frac{(-1+20\beta^2)}{108}\frac{m_{\varLambda}^3}{\beta^3}\approx \frac{5}{27}\frac{m_{\varLambda}^3}{\beta}.
\]
Using $p$ and $q$ we can evaluate D given which gives
\begin{equation}
D=-\frac{1}{46656}\left(\frac{m_{\varLambda}}{\beta}\right)^6+\frac{25}{2916}\frac{m_{\varLambda}^6}{\beta^2}>0
\end{equation}
which is positive, at least for 
$1<\beta<\sqrt{3}$ ($\beta=1$ corresponds to the maximum value of the horizon $r_3$ whereas
$\beta =\sqrt{3}$ to the maximum value of $r_2$). 
Thus we have now $D>0$ and $p<0$ which is case ii) in appendix A. 
The parameter $R$ and the auxiliary angle come out to be $R=\frac{m_{\varLambda}}{6\beta}$ and $\cosh\phi=20\beta^2$ which
for $\beta=1$ is $\phi=3.68$. 
The real solution is
\begin{equation} \label{real}
T_*=R\left(1-2\cosh (\phi/3)\right)
\end{equation}
which is never positive.
Indeed, the first step to get a real {\it positive} solution is to return to case i) in the appendix A which
requires $D <0$. Hence, putting $D(\beta)=0$ gives $\beta = 1/\sqrt{10}$ in agreement with
$2r_s(M_{\rm max}^*) = r_\varLambda/\sqrt{10}$. In the context of GUP we therefore have a minimum and maximum horizon defined by
\begin{equation} \label{maxhor}
r_{*c}^{\rm min}= l_{\rm pl}, \quad r_{*c}^{\rm max}=2r_s(M_{\rm max}^*)=\frac{1}{\sqrt{10}}r_\varLambda 
\end{equation}

\section{An independent source of information on $T_{\rm min}$ and $T_{\rm max}$}
In 1966 Andrei Sakharov found a maximum temperature of black body 
radiation to be of the order of Planck mass \cite{sakharov} 
\begin{equation} \label{sakharov}
{\cal T}_{\rm max}^{\rm Sakharov} \approx m_{\rm pl}
\end{equation}
He based his results on very general arguments. 
This result  is confirmed in equation (\ref{eq:maxumus}) which is of the same order
of magnitude as (\ref{sakharov}).
Sakharov's result bears a certain importance. Combined with Hawking's formula for black hole
evaporation $T=1/(8\pi G_N M)$, it implies independently of GUP the existence 
of a black hole remnant of the order
of Planck mass. Indeed, the value of the maximal temperature is $\sim 10^{32}\,\, K$
and has only a physical relevance in black hole evaporation. We can show yet a third way, to establish this
important result. This method is then also suitable to include $\Lambda$.
The $-g_{00}$ component of the metric should be 
positive definite (see chapter 84 in \cite{LL3} for a general discussion). 
We can regard also the mass $M$ entering the Schwarzschild metric
as energy which, in turn, can be replaced by energy density $\rho$ i.e. $0 < -g_{00}=1-\frac{2G M}{R}=1-(8\pi/3)G \rho R^2$. 
Hence $\rho < \frac{3}{8\pi}\frac{1}{G R^2}$. Using the Stefan-Boltzmann law $\rho=\sigma T^4$ gives \cite{massa1} $T^4 < \frac{3}{8\pi}\frac{1}{\sigma G R^2}$. Finally, to get rid of the radius $R$ we employ the
quantum mechanical result for black body radiation, $R > 1/T$ \cite{Bekenstein, massa1, massa2}. 
The maximal temperature obtained this way, namely $T < {\cal T}_{\rm max}=\sqrt{\frac{45}{8\pi^3}}m_{\rm pl}$
 is of the same order of magnitude as ${\cal T}_{\rm max}$ in equation (\ref{eq:maxumus}).
Repeating the same steps $\varLambda \neq 0$ i.e. for the Schwarzschild-de Sitter metric
we can write
\[
0 < \rho < \frac{3}{8\pi}\frac{m_{\rm pl}^2}{R^2}-\frac{1}{3}
\frac{3}{8\pi}m^2_{\rm pl}m^2_\varLambda < {\cal F}=\frac{3}{8\pi}m_{\rm pl}^2 T^2-\frac{1}{3}
\frac{3}{8\pi}m^2_{\rm pl}
m_{\varLambda}^{2}
\]
where we used again $ R > 1/T$.
One of these inequality, $\rho < {\cal F}$, gives us back ${\cal T}_{\rm max}$ with small
correction due to $\varLambda$. The other one, $0< {\cal F}$, can be translated into
${\cal T}_{\rm min}$ such that in the end we get $\frac{1}{\sqrt{3}}m_{\varLambda}
={\cal T}_{\rm min} < T < {\cal T}_{\rm max} \sim m_{\rm pl}$ confirming the existence of a minimal and maximal temperature in a different way.

\section{Gravitational red-shift for the temperature with the presence of $\varLambda$.}
In this section we study yet another consequence of $\varLambda$ in
the measurement of temperature at a distance $r_2^*$.
This effect manifests itself in the 
gravitational red-shift of thermal and electromagnetic radiation.
To compare the two cases, we work first on the electromagnetic part.
We follow here in parts \cite{carroll}. In equation (\ref{eq:red-shiftfactor}) we obtained the red-shift factor 
$V=\sqrt{-K^{\sigma}K_{\sigma}}$ in the Schwarzschild-de Sitter metric 
which we can use in the  standard relation of red-shift for electromagnetic 
wavelength \cite{carroll} for static observers:
\begin{equation}    \label{eq:rsfwwl}
\lambda_2=\frac{V_2}{V_1}\lambda_1={\cal Z}_{\rm grav}\lambda_1
\end{equation}
where a photon with wavelength $\lambda_1$ has been emitted at a distance $r_1^*$ and detected
as $\lambda_2$ at $r_2^*$.
It can be demonstrated that at the Killing horizons \cite{carroll}
the red-shift factor is zero, i.e., we have $V(r\approx 2r_s)=V(r\approx\sqrt{3}r_\varLambda)=0$
 which is also evident from the explicit form
\begin{equation}   \label{eq:Possbs}
\displaystyle{\lambda_2={\cal Z}_{\rm grav}(r_2^*, r_1^*)\lambda_1=
\left(\sqrt{\frac{1-\frac{2r_s}{r^*_2}-\frac{1}{3}\frac{r_2^{*2}}{r_\varLambda^2}}{1
-\frac{2r_s}{r^*_1}-\frac{1}{3}\frac{r_1^{*2}}{r_\varLambda^2}}}\right)\lambda_1}
\end{equation}
For a fixed mass satisfying the condition $r_s<<r_\varLambda$, 
a photon emitted by a static observer 1 will be observed by static observer 2 
at a distance $r\approx\sqrt{3}r_\varLambda$ with a wavelength $\lambda_2$ given by
\begin{equation}   \label{eq:bshif23}
\lambda_2=\frac{V_2(r\approx\sqrt{3}r_\varLambda)}{V_1}\lambda_1\approx0
\end{equation}
If this is true, there must be some distance $r_0$
after which the wavelength 
$\lambda_2$ begins to decrease in contrast
to what happens in the case $\varLambda=0$ where
${\cal Z}_{\rm grav} \to 1/V_1$ as $ r_2^* $ becomes large.
Suppose a photon is emitted at $r_1^*$ and detected as $\lambda_2$ at
$r_2^* \to r_0$. If the photon is emitted at the same distance, but detected as
$\lambda_2^{\prime}$ at
$r_2^* > r_0$, then $\lambda_2^{\prime}$ 
is blue-shifted as compared to $\lambda_2$.
The distance $r_0$ can be found if we consider the function \cite{we}
\begin{equation}   \label{eq:thismar}
e^{\nu(r)}=1-\frac{2r_s}{r}-\frac{1}{3}\frac{r^2}{r_\varLambda^2}
\end{equation}
It has a local maximum at $r_0=(3r_sr_\varLambda^2)^{1/3}$ where $r_0$ 
coincides with the distance after which 
a test body has no bound orbits in the Schwarzschild-de Sitter metric \cite{we2} (see also equation
(\ref{lmax})). 
We can summarize the results by saying that $\lambda_2$ increases as $r$ increases up to the value $r_0$ whereas $\lambda_2$ decreases as $r$ increases starting from $r_0$. The maximal red-shift experienced by electromagnetic waves is
\begin{equation}
{\cal Z}_{\rm grav}^{\rm max} =\frac{\sqrt{1-2\left(\frac{r_s}{r_\varLambda}\right)^{2/3}}}{V_1}
\end{equation}
after which $\lambda_2$ becomes smaller.
Worth noting is the fact that $r_0$ is of astrophysical order of magnitude
as it is a combination of a large and a small distance \cite{we2}.

A similar procedure can be repeated for thermal radiation \cite{carroll} starting with
\begin{equation}
T_2=\frac{V_1}{V_2}T_1=\frac{V_1}{V_2}\frac{a_1}{2\pi}
\end{equation}
where $a_1$ is the invariant acceleration \cite{carroll}.
The observed temperature at a given distance $r_2^*$ is given by
\begin{equation}
T(r_2^*)=\lim_{r_1^*\to 2r_s}\frac{V_1a_1}{2\pi V_2}=
\frac{1}{V_2}\frac{\kappa}{2\pi}
=\frac{\kappa/2\pi}{\sqrt{1-\frac{2r_s}{r^*_2}-\frac{1}{3}\frac{r^{*2}_2}{r_\varLambda^2}}} 
\end{equation}
As $r^*_2\to(3r_sr_\varLambda^2)^{1/3}$, we obtain approximately
\begin{equation}
T(r_2^* \to r_0)\approx\frac{\kappa/2\pi}{\sqrt{1-2\left(\frac{r_s}{r_\varLambda}\right)^{2/3}}}
\end{equation}
which corresponds to a minimal temperature at a finite distance. After $r_0$ the temperature increases with $T(r_2^*)\to\infty$ as $r^*_2\to\sqrt{3}r_\varLambda$ which shows the difference with the case of the photon's wavelength.
As before we summarize this result by saying that $T_2$ decreases as $r$ increases up to the value $r_0$ while $T_2$ increases starting from $r_0$. 

\section{Conclusions}
As compared to the standard Hawking radiation formula $T(M)=m_{\rm pl}^2/(8\pi M)$, the
results from the Generalized Uncertainty Principle \cite{adler1, adler2}, with $\varLambda=0$, 
give a slightly different picture which agrees, as far as the existence of some minimal/maximal
physical quantities (mass, length etc.) is concerned, with expectations from quantum gravity:
\begin{itemize}
\item[1.] There exists a black hole remnant with a mass $M_{\rm min} \sim m_{\rm pl}$
corresponding to a maximal temperature $T_{\rm max} \sim m_{\rm pl}$.
\item[2.] There exists a minimum length (minimum horizon) $r_{*c}^{\rm min} \sim l_{\rm pl}$.
\item[3.] For large masses as compared to $M_{\rm min}$, $T(M)$ goes over to the standard Hawking
formula. For masses close to $M_{\rm min}$ (equivalently, for the horizon close to $r_{*c}^{\rm min}$), $T(M)$
gets deformed.
\end{itemize}

With the inclusion of $\varLambda$, Einstein's gravity becomes a two-scale theory which
in our universe has a hierarchical structure: $m_{\rm pl} \gg m_\varLambda$, $r_\varLambda \gg 
r_s \gg l_{\rm pl}$ etc.. It is known that $ \varLambda$ has a dual effect in the sense
that if a quantity is restricted by some maximal (minimal) 
value connected to the Newtonian constant $G$, $\varLambda$ has the opposite effect,
i.e., it introduces a minimal (maximal) restriction.  Two such examples have been explicitly
given in equations (\ref{lmin}) and (\ref{lmax}) which also demonstrate the fact that
$\varLambda$ has local effects \cite{wetherest}. It is then not unreasonable to ask
if a Generalized Uncertainty Principle with $\varLambda$ displays dual effects to
the points 1-3 above. To this end, we formulated an uncertainty relation with $\varLambda$ 
along the same lines of arguments used in the standard GUP case. We applied it to
Hawking radiation and found that (in doing so we emphasized certain aspects and neglected other
\cite{Others} which would not change our results):
\begin{itemize}
\item[4.] The results 1-3 from above hold.
\item[5.] There exists a maximum mass due to $\varLambda$ whose value is
$M_{\rm max} \sim (m_{\rm pl}/m_\varLambda)m_\varLambda$ corresponding
a minimum temperature $T_{\rm min} \sim m_\varLambda$.
We obtain the same result in section five
by looking into black body radiation. This again confirms that the new GUP relation
is consistent.

\item[6.] There exists a maximum length (at least in in black hole radiation context)
$r_{*c}^{\rm max} \sim  r_\varLambda/3= 1/3\sqrt{\varLambda}$. Beyond this value
the GUP equation as applied to black hole evaporation does not have any solution.
\item[7.] For intermediate masses the $T(M)$ dispersion relation derived via GUP
goes over to the standard relation (\ref{eq:why?}) derived via the surface gravity.
To put it in a in different words,  
the fact
that $\gamma=5/9$ comes out of order of unity  tells us that the GUP relation with $\varLambda$ is correct.
For even smaller masses this goes over to the Hawking formula $T \propto 1/M$
which in turn gets replaced by the the deformed relation for masses close
to $m_{\rm pl}$ (see point 3 above). For masses close to $M_{\rm max}$ (equivalently,
for the horizon close $r_{*c}^{\rm max}$) $T(M)$ also gets modified as compared to (\ref{eq:why?}).
\end{itemize}
Some of the above results find an independent confirmation. Notable is first of all, the 
paper by Sakharov \cite{sakharov} who in 1966 derived the maximal temperature being of the order
of Planck mass.  His line of arguments are different from GUP (indeed, in 1966
Hawking radiation has not been discovered yet). In section 5 we also showed that
the maximal and minimal temperature can be confirmed from the 
Schwarzschild-de Sitter metric by using the the Stefan-Boltzmann law and
a quantum mechanical restriction on $R$, i.e., $R > 1/T$. 

It is clear that in 1966 a temperature of the order of Planck mass which in units of Kelvin
is $10^{32}K$ was theoretically inaccessible in the sense that no
available theory produced such a temperature. 
Evidently, extreme situations are asked here for.
It is only with the advent of Hawking radiation that
$T_{\rm max}$ makes phenomenologically sense. Quite similarly $T_{\rm min} \sim 10^{-29}K$
(we used the fact that today the preferable value of $\varLambda$ is given by
$\rho_{\rm vac}=\varLambda/(8\pi G)\approx 0.7 \rho_{\rm crit}$) requires equally an extreme
situation and is a temperature which appears only
in connection with black hole evaporation. Sakharov derived his $T_{\rm max}$ as a maximum
temperature of black body radiation (we can apply it to black holes since the spectrum of the latter is
the one of black body radiation). It appears that $T_{\rm min}$ might enjoy also a
broader interpretation as the minimal temperature which can be reached in nature, at least
in principle.

If we look back, the speculations that the Planck length is the smallest length in
nature were based on purely dimensional analysis.
GUP confirms this expectation when applied to Hawking radiation.
On the other hand GUP predicts also a maximal length of the order
$r_\varLambda$, again the context of black holes. 
Quantum gravity effects become not only important at $l_{\rm pl}$, but
evidently also at $r_\varLambda$.
Such a result could have been 
also guessed (as opposed to explicitly demonstrated as e.g. in GUP)
on the basis of scale analysis and therefore we might {\it speculate} that
the maximal length, as its minimal counterpart,  has a broader 
meaning as a maximally possible length in nature. If so,
there should be interesting consequences for cosmology in our universe which
is right now dominated by $\varLambda$. This is to say, 
$r_{\varLambda}=\frac{1}{\sqrt{\varLambda}}=
      \frac{1}{\sqrt{3}}\left(\frac{\rho_{\rm vac}}{\rho_{\rm crit}}\right)^{-1/2}
      H_0^{-1}$
which means that the Hubble radius is almost $r_\varLambda$ at the present epoch.

The example with the Hubble radius is also interesting from the perspective of cosmological
coincidences. Not only the Hubble radius is dominated by $\varLambda$, but it is also
worth mentioning that the maximal mass ( see eq.(\ref{eq:mq02})), which we found, is also
close to the mass of the universe.  Such coincidences might be of interest in the framework
of different theories \cite{smolin1, barrow}.

Finally, it makes sense to compare our findings with similar results obtained elsewhere. The temperature
$T_{\rm GH}=(\Lambda /3)^{1/2}/(2\pi)$ is known as Gibbon-Hawking temperature \cite{Wald}. The order
of magnitude of the Gibbon-Hawking temperature corresponds to the minimum temperature found in the present paper.
An inertial observer in a de Sitter universe is immersed in a thermal bath at the Gibbon-Hawking temperature.
This led the authors of \cite{Dalal} to the estimate of a minimum temperature and a maximum mass.
Briefly their argument goes as follows. The change of the black hole mass is $\dot{M} \propto (T^4 _{\rm GH}
-T^4)$. Insisting on $\dot{M} \le 0$ leads to the results that the minimum temperature is given by the
Gibbon-Hawking expression which, up to order of magnitude, agrees with our result. Postulating $\dot{M}=0$
gives $M_{\rm crit}=(1/4G)\sqrt{3/\Lambda}$ which again agrees with maximum mass we found via the
Generalized Uncertainty Principle. The arguments used in \cite{Dalal} (see also \cite{Baez})
are different from the derivation used
in the present paper. They lead, however, to the same results. This confirms the Generalized Uncertainty Principle 
with $\Lambda$ which was our starting point and corroborates our results derived from it.
  
\renewcommand{\theequation}{A-\arabic{equation}}
% redefine the command that creates the equation no.
\setcounter{equation}{0}  % reset counter 

\section*{Appendix A: General solution of a third-order polynomial} 
In the present paper we have been using many times the parametric solution
of a third order polynomial. For the reader's convenience and to
set up general definition, we outline below the three different cases
of the zeroes of the third order polynomial \cite{bronstein}.
The standard equation of a third-order polynomial is
\begin{equation}   \label{eq:third-order}
x^3+rx^2+sx+t=0
\end{equation}
The reduced form of the third-order equation (\ref{eq:third-order}), requires the change of variable
\begin{equation}   \label{eq:covto}
y\equiv x+\frac{r}{3}
\end{equation}
such that the reduced form is given by
\begin{equation}   \label{eq:rt}
y^3+py+q=0
\end{equation}
The corresponding coefficients read
\[
p=s-\frac{r^2}{3},\;\;\;\;q=\frac{2}{27}r^3-\frac{rs}{3}+t
\]  
It is necessary to establish some classification criteria for the 
solutions of the reduced third-order equation. These criteria are based on a parameter $D$ defined by
\[
D\equiv \left(\frac{p}{3}\right)^3+\left(\frac{q}{2}\right)^2
\]
A second important parameter $R$ entering the parametric solutions is
\[
R\equiv {\rm sign}(q)\sqrt{\frac{\vert p\vert}{3}}
\]
The solutions are parametrized by an auxiliary angle $\phi$ whose exact definition depends on the signs of $p$ and $D$.
We distinguish three cases: \\\\

Case i) $p<0$, $D\leqslant 0$\\

In this case, the auxiliary angle is defined as:
\[
\cos\phi\equiv \frac{q}{2R^3}
\]
with the corresponding solutions all real and given by
\[
y_1=-2R\cos\frac{\phi}{3},\quad 
y_2=-2R\cos\left(\frac{\phi}{3}+\frac{2\pi}{3}\right),\quad
y_3=-2R\cos\left(\frac{\phi}{3}+\frac{4\pi}{3}\right) 
\]

Case ii) $p<0$, $D>0$\\

In this case, the auxiliary angle is 
\[
\cosh\phi\equiv \frac{q}{2R^3}
\]
and the corresponding solutions are:
\[
y_1=-2R\cosh\frac{\phi}{3},\quad
y_2=R\cosh\frac{\phi}{3}+i\sqrt{3}R\sinh\frac{\phi}{3},\quad
y_3=y_2^*=R\cosh\frac{\phi}{3}-i\sqrt{3}R\sinh\frac{\phi}{3}
\]

Case iii) $p>0$, $D>0$\\

In this section, we define the auxiliary angle to be:
\[
\sinh\phi\equiv \frac{q}{2R^3}
\]

with the explicit solutions:
\[
y_1=-2R\sinh\frac{\phi}{3},\quad
y_2=R\sinh\frac{\phi}{3}+i\sqrt{3}R\cosh\frac{\phi}{3},\quad
y_3=y_2^*=R\sinh\frac{\phi}{3}-i\sqrt{3}R\cosh\frac{\phi}{3}
\]

\renewcommand{\theequation}{B-\arabic{equation}}
% redefine the command that creates the equation no.
\setcounter{equation}{0}  % reset counter 

\section*{Appendix B: Invariant scalar four-acceleration for static observers}
In agreement with \cite{Massadough2}, a typical static observer
will have a four-velocity given by (we use here natural units)
\[
u^\alpha=\frac{dx^\alpha}{d\tau}=(u^0,0,0,0),\quad u^0=\frac{dt}{d\tau}=(-g_{00})^{1/2}
\]
The observer's proper four-acceleration components will be
\[
a^\alpha=\frac{du^\alpha}{d\tau}=u^\nu u^\alpha_{\;\;;\nu}
\]
Explicitly, this equation reads $a^\alpha=\left(u^\alpha_{\;\;,\nu}+\Gamma^\alpha_{\sigma \nu}u^\sigma\right)u^\nu$
 or $a^\alpha=\left(u^\alpha_{\;\;,0}+\Gamma^\alpha_{0 0}u^0\right)u^0$ for a local static observer with $u^j=0$.
In local static coordinates the condition $u^0_{\;\;,0}=0$ is satisfied. Thus we obtain
the simple expression $a^\alpha=\Gamma^\alpha_{0 0}(-g_{00})^{-1}$ where  we used the fact that $u^0u^0=(-g_{00})^{-1}$. 
For the explicit calculation of $a^\alpha$, 
it is necessary to evaluate the Christoffel connection
and use the fact that the metric is static, i.e., $g_{\mu \nu},_0=0$. Then we obtain
\[
a^\alpha=\frac{1}{2}g^{\alpha \mu}g_{00,\mu}g^{-1}_{00}
\]
In the Schwarzschild-de Sitter metric all non-radial components vanish $a^0=a^\theta=a^\phi=0$ and the only surviving component is \cite{Massadough2}
\begin{equation}   \label{eq:ra}
a^r=\frac{1}{2}g^{rr}g_{00,r}g_{00}^{-1}
\end{equation}
The Schwarzschild-de Sitter metric is given by
\begin{equation}   \label{eq:1stmetric}
g^{rr}=1-\frac{2r_s}{r}-\frac{r^2}{3r_\varLambda^2}=-g_{00},\quad
g^{00}=-g_{rr}=g_{00}^{-1}
\end{equation}
Therefore we can write
\[
g_{00,r}=2\left(-\frac{r_s}{r^2}+\frac{1}{3}\frac{r}{r_\varLambda^2}\right)
\]
Replacing (\ref{eq:1stmetric}) in (\ref{eq:ra}) we arrive at
\[
a^r=-\frac{1}{2}g_{00,r}=\frac{r_s}{r^2}-\frac{1}{3}\frac{r}{r_\varLambda^2}
\]
The invariant acceleration can be calculated to be 
\[
a=\sqrt{g_{\mu \nu}a^\mu a^\nu}=\frac{1}{2}\sqrt{g_{rr}}(g_{00}g_{rr})^{-1}\frac{dg_{00}}{dr}
\]
where we made use of $g_{00}g_{rr}=-1$.

\end{document}